\documentclass[%
 aip,
 jmp,%
 amsmath,amssymb,
 reprint,%
]{revtex4-1}

\usepackage{graphicx}
\usepackage{dcolumn}
\usepackage{bm}
\usepackage{subfigure}
\usepackage{color} 

\begin{document}

\preprint{AIP/123-QED}

\title[]{On the Anomalous Diffusion in Nonisothermal plasma}

\author{A. A. Rukhadze}
\email{rukh@fpl.gpi.ru}
\altaffiliation{A. A. Rukhadze and S. P. Sadykova contributed equally to this work.}
\affiliation{Prokhorov General Physics Institute,
Russian Academy of Sciences, Vavilov Str., 38., Moscow, 119991, Russia}
\author{S. P. Sadykova}%
\altaffiliation{Electronic mail: \textcolor{blue}{Corresponding author - saltanat@physik.hu-berlin.de.}}
\affiliation{Humboldt-Universit\"at zu Berlin, Berlin, 12489, Germany, Newtonstr. 15
}%


\date{\today}

\begin{abstract}
In nonisothermal plasma at temperature $T_e>> T_i$ diffusion plays decisive role at conditions of smooth inhomogeneity when the inhomogeneity size is larger than the Debye radius by more than $\sqrt{T_e/T_i}$ times. When the inhomogeneity is rather abrupt and the condition is violated, then during the spreading process the Maxwellian relaxation of ion charges becomes significant. Here, we consider these two phenomena together and refer to the anomalous character of diffusion, i.e. anomalous diffusion. %

\end{abstract}

\maketitle

\section{\label{sec:Int}Introduction}

One can talk of plasma as a continous medium when its size (and the inhomogenuity size) exceeds significantly the Debye radius. In a case of nonisothermal plasma $T_e>>T_i$  Debye radius is determined through the ion temperature and is equal to $v_{Ti} /\omega_{Li}$, where $v_{Ti}=\sqrt{T_i/M}$ is the ion thermal velocity and $\omega_{Li}=\sqrt{4\pi {q_i}^2 n_{i0}/M}$ is the ion Langmur frequency with $q_i=Ze$ being the ion charge, $n_{i0}$ - equillibrium ion density ($e n_{e0}=q_i n_{i0}$), here $e$ - electron charge. Here and below we use the CGS unit system. We write for general case but further on we present the results of our calculations only for the Hydrogen plasma Z=1. \\
\indent The diffusion process is characterized through the mean square displacement dependence on time:
\begin{equation}
\label{111}
<r^2(t)>=\frac{\int r^2 n(\vec r, t) d \vec r }{\int n(\vec r, t) d \vec r}\sim t^\gamma,
\end{equation}
If $\gamma=1$ then diffusion is termed as normal, in all other cases one refers to the anomalous character of diffusion: $\gamma>1$ - superdiffusion, $\gamma<1$- subdiffusion.\\
\indent  When the size of innhomogeneity of ionic component perturbation is larger than the Debye radius by more than $\sqrt{T_e/T_i}$ times, then spreading of the ionic component perturbation has a diffusive character and is defined by the time of the normal ambipolar diffusion \cite{Princpls}. If it is less then the Maxwellian relaxation of ionic charges becomes dominant during the spreading of  inhomogenuity \cite{Lasers,Waves}. \\
\indent We will introduce a reader below into the theory of spreading of perturbation in nonisothermal plasma taking into account both mentioned phenomena and show the anomalous character of diffusion. 

\section{Anomalous diffusion model}
First of all, we will consider the main equations of the studied phenomenon and its constraints.    \\
\indent The main constraints for the considered below equations are:
\begin{equation}
\label{1}
\omega \nu_i >> \omega^2;\: k^2 v_{Te}^2 >>\omega \nu_e;\: \nu_e >> k v_{Te}, \omega;\: \nu_i >> k v_{Ti}, \omega
\end{equation} 
Here $\nu_e$, $\nu_i$ are electron-atom and ion-atom collision frequencies, $v_{Te}=\sqrt{T_e/m}$ - electron thermal velocity with $m$ being the electron mass. Nonisothermal plasma can be only the weakly ionized plasma, in which the dominant collisions are charged particle-atom, -molecule collisions, $\omega \propto 1/\tau $, where $\tau$ is the characteristic time of spreading and $k \propto 1/L $, where $L$ is the characteristic size of spreading.   \\
\begin{itemize}
 \item[1] Equation of motion of electrons and ions has the following view: 
\begin{equation}
\label{2}
\frac{q_i \vec E}{M}- \nu_i \vec v_i =0, \: q_e\vec E -\frac{T_e}{n_e}\Delta n_e =0,
\end{equation}
Here $q_e=e$, $q_i=Z e$ are the electron and ion charges, $Z \leq |q_i/q| $, $\vec E$ are the self-consistent field,  $\vec v_i$, $\vec v_e$ are ion and electron velocities, $T_i=0$, $T_e=const$, $n_i$, $n_e$ are the ion and electron densities.
\item[2] Field equation (Poisson equation)
 \begin{equation}
\label{3}
div \vec E= 4\pi e(n_e-Zn_i)
\end{equation}
\item[3] Continuity equation of the ion component
 \begin{equation}
\label{4}
\frac{\partial n_i}{\partial t}+ div \: n_i \vec v_i=0
\end{equation}
\item[4] Equilibrium state satisfying the stationary equations (\ref{2})-(\ref{4}) at $\vec E_0=0$ takes the following view:
\begin{equation}
\label{5}
q_e n_{e0}=q_i n_{i0}=Ze n_{i0}=Const
\end{equation}
  Let's consider the small density perturbations :
\begin{equation}
\label{6}
n_{e}=n_{e0}+ \delta n_e, \: n_i=n_{i0}+ \delta n_i,
\end{equation}
where $n_{\alpha0}$ - denity of the homogenious background. These small density perturbations lead to the emergence of the weak self-consistent field $\vec E$.\\
\end{itemize}
\indent System of equations (\ref{2})-(\ref{4}) at (\ref{5}) for small perturbations (\ref{6}) can be deduced to one equation for perturbation $\delta n_i$. 
The rest magnitudes can be expressed through $\delta n_i$. Equation for $\delta n_i$ has the following view:
 \begin{equation}
\label{7}
({\omega_{Li}}^2-{v_S}^2\Delta)\frac{\partial \delta n_i}{\partial t}-\frac{{v_S}^2}{\nu_i}{\omega_{Li}}^2 \Delta \delta n_i=0.
\end{equation}
Here $\omega_{Li}$ is the equlibrium  Langmuir ion frequency, ${v_S}^2=ZT_e/M$ is the squared ion sound velocity and $\nu_i=Const$. Eq. (\ref{7}) represents the partial differential equation of the first order with the constant coefficients.\\
\indent Let us formulate the Koshi problem for Eq. (\ref{7}), i.e. we consider infinite plasma and set the following initial condition:  
\begin{equation}
\label{8}
\delta n_i(\vec r,0)=f(\vec r),
\end{equation}
and take into account that $ lim _{r \to\infty } \delta n_i=0$. Using the Fourier transform 
\begin{equation}
\label{9}
\begin{gathered}
 \delta n_i(\vec r,t)=\int d \vec k \delta n_i (\vec k, t) \exp(i \vec k \cdot \vec r),  \hfill\\ f(\vec k)=\frac{1}{(2\pi)^3}\int d \vec r f (\vec r) \exp(-i \vec k \cdot \vec r),
\end{gathered}
\end{equation}
\indent Eq. (\ref{7}) will be reduced to the ordinary differential equation of the first time order.
 \begin{equation}
\label{10}
\frac{\partial\delta n_i(\vec k,t)}{\partial t}= -\frac{k^2{\omega_{Li}}^2{v_S}^2/\nu_i}{{\omega_{Li}}^2+k^2{v_S}^2}\delta n_i(\vec k,t).
\end{equation}
We obtain the solution of Eq. (\ref{10}) with the initial condition
 \begin{equation*}
\label{11}
\delta n_i(\vec k,0)= f(\vec k),
\end{equation*}
as following:
\begin{equation}
\label{12}
\delta n_i(\vec k,t)= f(\vec k)\exp\left(-t\frac{k^2{\omega_{Li}}^2{v_S}^2/\nu_i}{{\omega_{Li}}^2+k^2{v_S}^2}\right).
\end{equation}
Making the inverse Fourier transform (\ref{9}), we can find
 \begin{equation}
\label{13}
\delta n_i(\vec r,t)=\int d\vec k f(\vec k) \exp\left(i \vec k \cdot \vec r -t\frac{k^2{\omega_{Li}}^2{v_S}^2/\nu_i}{{\omega_{Li}}^2+k^2{v_S}^2}\right).
\end{equation}
Let us analyse the limitting cases of Eq. (\ref{13}). 
\begin{itemize}
\item[A]. In a case of rare plasma, when  ${\omega_{Li}}\to 0$:
 \begin{eqnarray}
\label{14}
\delta n_i(\vec r,t)=&&\int d\vec k f(\vec k) \exp\left(i \vec k \cdot \vec r -t\frac{{\omega_{Li}}^2}{{\nu_i}}\right)\nonumber \\
&&=f(\vec r)\exp\left( -t\frac{{\omega_{Li}}^2}{{\nu_i}}\right).
\end{eqnarray}
This solution corresponds to the Maxwellian temporary relaxation of the charge density perturbation. 
\item[B]. In a case of dense plasma, when  ${\omega_{Li}}^2 >> k^2 {v_S}^2$:
\begin{equation}
\label{15}
\delta n_i(\vec r,t)=\int d\vec k f(\vec k) \exp\left(i \vec k \cdot \vec r -t\frac{k^2 {v_S}^2}{\nu_i}\right).
\end{equation}
The integral can be easily solved at Eq. (\ref{8}), when $f(\vec r)=N_{i0}\delta (\vec r)$ with  the corresponding Fourier transform $f(\vec k)=\frac{N_{i0}}{(2\pi)^3}$, where $N_{i0}$ is the total number of perturbed particles at $r=0$. The result is well known \cite{Lan}:
\begin{equation}
\label{16}
\delta n_i(\vec r,t)=\frac{N_{i0}}{8(\pi D t)^{3/2}} \exp\left(-r^2 /4Dt \right ),
\end{equation}
where $D={v_S}^2/\nu_i$ is the coefficient of ambipolar difusion and Eq. (\ref{16}) describes the diffusive spreading of the localized ion density perturbation ($\delta$ - functional).\\
\indent Let us determine the $<r^2(t)>$ Eq. (\ref{111}) for the ambipolar diffusion (\ref{16}):
\begin{equation}
\label{112}
<r^2(t)>=6D t,
\end{equation}
here in Eq. (\ref{111}) we used $\delta n_i$ Eq. (\ref{17}). As one can see this corresponds to the normal diffusion.
\end{itemize}
\indent Let us choose the following two initial perturbations $f(\vec r)$, the first as representing the infinite wave:
\begin{equation}
\label{17}
f(\vec r)=n_{i0} \cos(\vec r \cdot \vec k_0) 
\end{equation}
with the Fourier transform $f(\vec k)=\frac{\delta(\vec k -\vec k_0)+\delta(\vec k +\vec k_0)}{2}$, here  $\vec k_0$ is the wave vector; 
and the second as repesenting the localized ion density perurbation mentioned above:
\begin{equation}
\label{700}
f(\vec r)=N_{i0}\delta (\vec r)
\end{equation}
with the Fourier transform $f(\vec k)=\frac{N_{i0}}{(2\pi)^3}$, what will be discussed later. We would lik to start with the condition (\ref{17}). \\
Then, the equation (\ref{14}) corresponding to the Maxwell relaxation  will turn into 
\begin{equation*}
\label{18}
\delta n_i(\vec r,t)=n_{i0} \exp \left(-\frac{\omega_{Li}^2 t}{\nu_i}\right)\cos(\vec r \cdot \vec k_0) 
\end{equation*}
or in dimensionless form
\begin{equation}
\label{19}
\frac{\delta n_i(\vec R,T)}{n_{i0}}= \exp \left(-\frac{\omega_{Li} T}{\nu_i}\right)\cos(\vec R \cdot \vec e_0), 
\end{equation}
here , $\vec k_0 r_D=\vec e_0$, $R=r/r_D$, $T=t\omega_{Li}$ - dimensionless time.\\
\indent The equation (\ref{15}) corresponding to the ambipolar diffusion will turn into 
\begin{equation*}
\label{20}
\delta n_i(\vec r,t)=n_{i0} \exp \left(-\frac{k_0^2 v_S^2 t}{\nu_i}\right) \cos(\vec r \cdot \vec k_0) 
\end{equation*}
or in dimensionless form
\begin{equation}
\label{21}
\frac{\delta n_i(\vec R,T)}{n_{i0}}= \exp \left(-\frac{k_0^2 D T}{\omega_{Li}}\right) \cos(\vec R \cdot \vec e_0), 
\end{equation}
here $\vec k_0$ is the same wave vector.\\
\indent Finally, the equation for anomalous diffusion (\ref{13}) will become
\begin{equation*}
\label{22}
\delta n_i(\vec r,t)=n_{i0}\exp\left(-t\frac{k_0^2 D{\omega_{Li}}^2{v_S}^2}{{\omega_{Li}}^2+k_0^2{v_S}^2}\right)\cos(\vec r \cdot \vec k_0).
\end{equation*}
or in dimensionless form
\begin{equation}
\label{23}
\frac{\delta n_i(\vec R,T)}{n_{i0}}=\exp\left(-t\frac{k_0^2 D{\omega_{Li}}{v_S}^2}{{\omega_{Li}}^2+k_0^2{v_S}^2}\right)\cos(\vec R \cdot \vec e_0).
\end{equation}
\indent As one can easily see in all cases, Eqs. (\ref{19})-(\ref{23}), the functions get separated by variables $T$ and $R$. As a result we can not define the mean square displacement. This means that at the given oscillatory initial perturbaion (\ref{17}) no normal diffusion occurs: the initial perturbation gets faded down  with the time evolution keeping its initial functional form in space - $\cos(\beta R)$, i.e. only the amplitude decreases. This behaviour is peculiar for the Maxwellian relaxation (\ref{14}) independent on the type of initial condition. However, when we take another initial condition ($\delta(\vec r)$ - functional) for the case of diffusive relaxation (\ref{15}) then we obtain the normal ambipolar diffusion (\ref{16}).  \\
\indent Finally, let us find the solution for the anomalous diffusion Eq. (\ref{13}) with the second initial condition Eq. (\ref{700}) as representing the localized ion charge perturbation. Having solved this equation and having made some simplifications, we have got the following:

\begin{eqnarray}
\label{701}
\frac{\delta n_i(r,t)}{N_{i0}/r_D^3} &=&\frac{r_D^3}{(2 \pi)^3}\int d\vec k \exp\left(i \vec k \cdot \vec r -t\frac{k^2{\omega_{Li}}^2{v_S}^2/\nu_i}{{\omega_{Li}}^2+k^2{v_S}^2}\right) \nonumber  \\
&=&\frac{r_D^3}{2 \pi^2} \int_0^{10\omega_{Li}/v_s} dk f(k) k  \frac{\sin(k r)}{r} \nonumber \\
&\times&\exp \left( - t \frac{k^2 v_s^2 \omega_{Li}^2/\nu_i}{\omega_{Li}^2+k^2 v_s^2}\right)
+\delta(\vec r) r_D^3 \exp \left( - t \frac{\omega_{Li}^2}{\nu_i} \right) \nonumber \\
&-&\frac{r_D^3}{2\pi^2} \exp \left(- t \frac{ \omega_{Li}^2}{\nu_i} \right)\int_0^{10\omega_{Li}/v_s}d k k  \frac{\sin(k r)}{r}
\end{eqnarray}  
\section{Results and Discussions}

\begin{figure*}
\subfigure
{\includegraphics[height=6.8 cm,width=7.7cm]{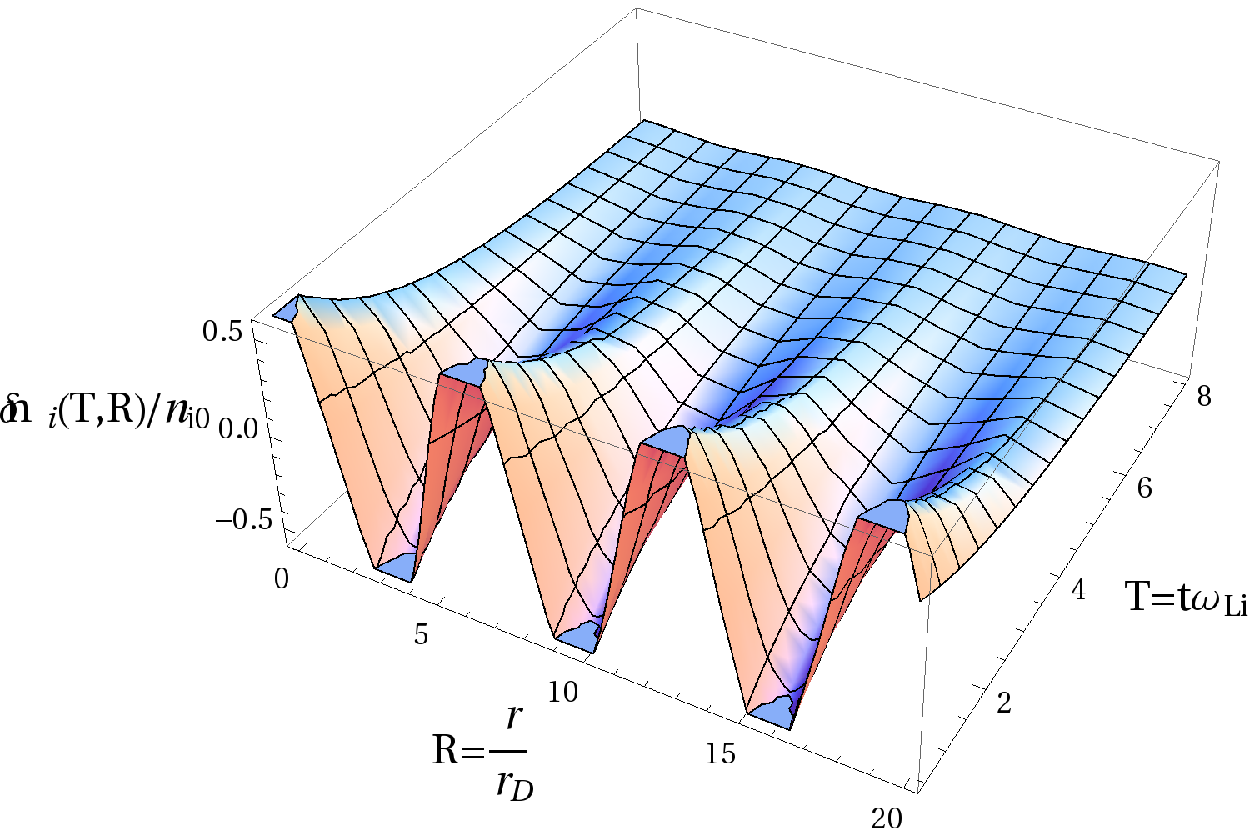}~a)}
\subfigure
{\includegraphics[height=6.8 cm,width=7.6cm]{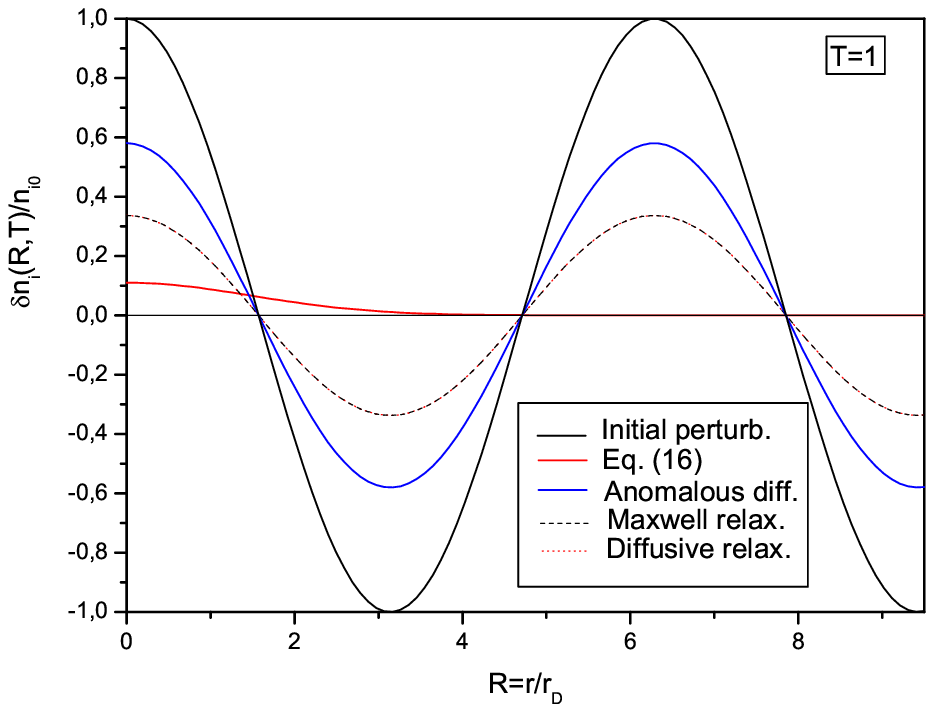}~b)}
\caption{a) The anomalous diffusion Eq. (\ref{23}) and b) comparison among the anomalous, Maxwellian relaxation Eq. (\ref{19}), ambipolar diffusive relaxation Eq. (\ref{21}) and the localized ion density perturbation relaxation Eq. (\ref{16}) at $t=1/\omega_{Li}$ is shown. Here, $\vec e_0 || \vec R$, $T_e=10$ eV, $n_{i0}=10^{12} cm^{-3}$, $\omega_{Li}=1.3 \cdot 10^{9} s^{-1} $, $\nu_i=1.2 \cdot 10^{9} s^{-1}$, $n_0=3 \cdot 10^{16} cm^{-3}$ (atom density), $r_D=2.3 \cdot 10^{-5} m$.  }
\label{Fig:Anom}
\end{figure*}

\begin{figure*}
\subfigure
{\includegraphics[height=6.8 cm,width=7.7cm]{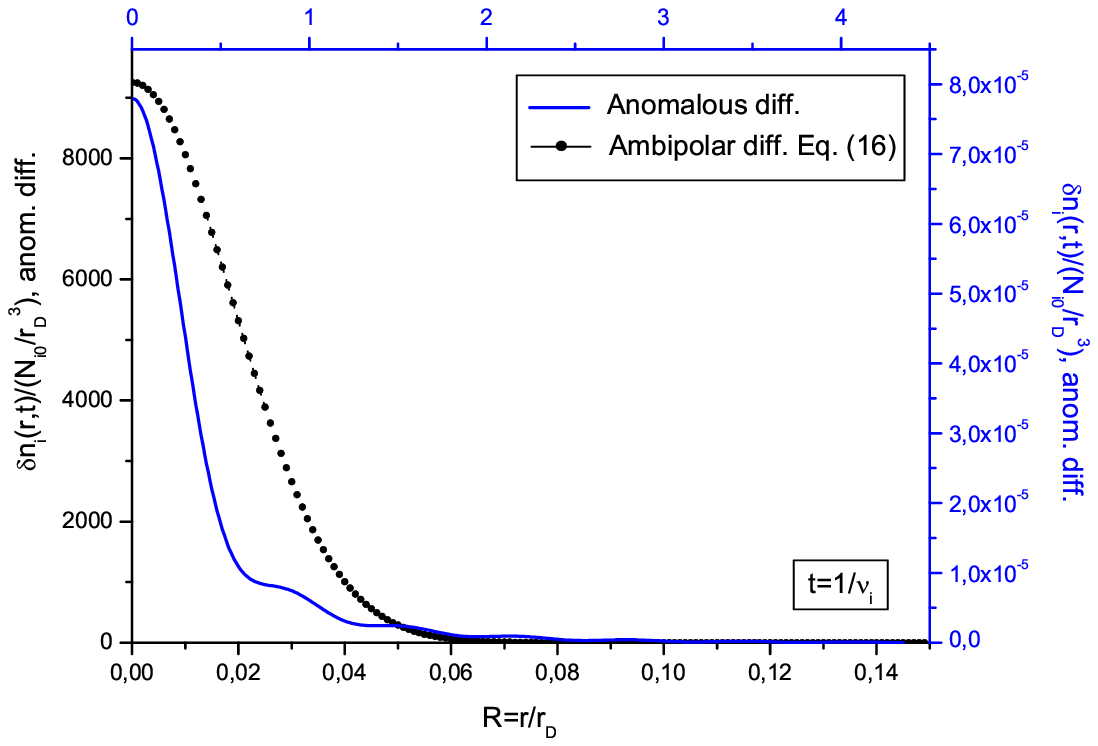}~a)}
\subfigure
{\includegraphics[height=6.8 cm,width=7.7cm]{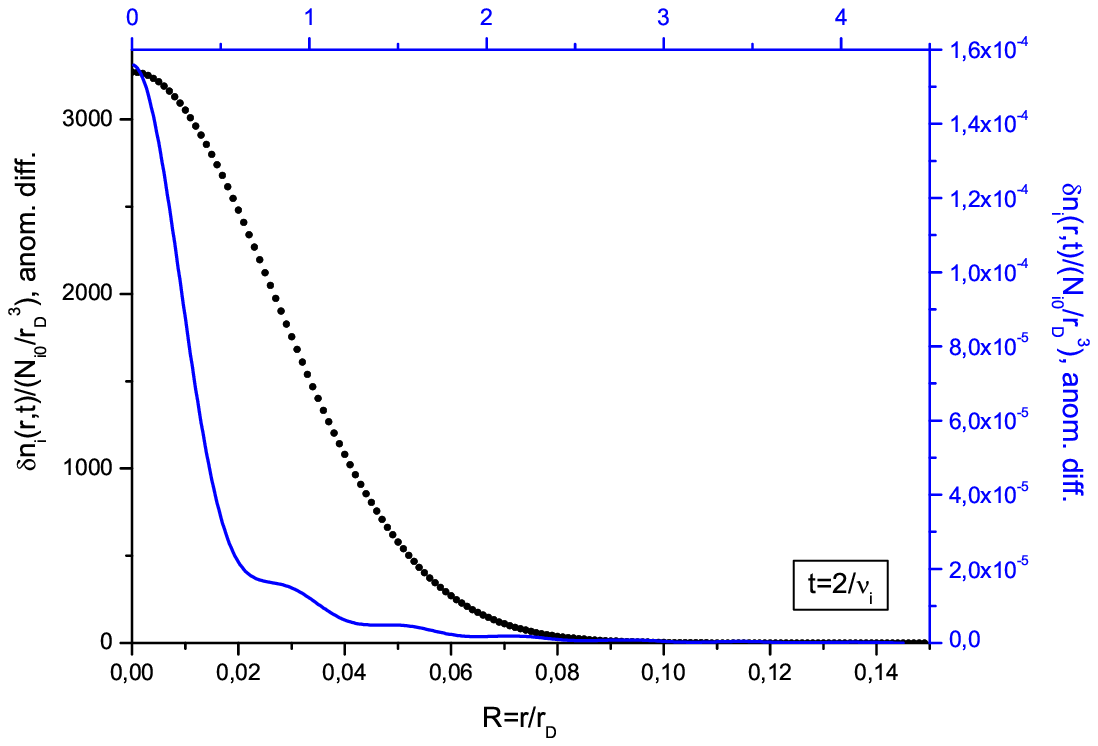}~b)}\\
\subfigure
{\includegraphics[height=6.8 cm,width=7.7cm]{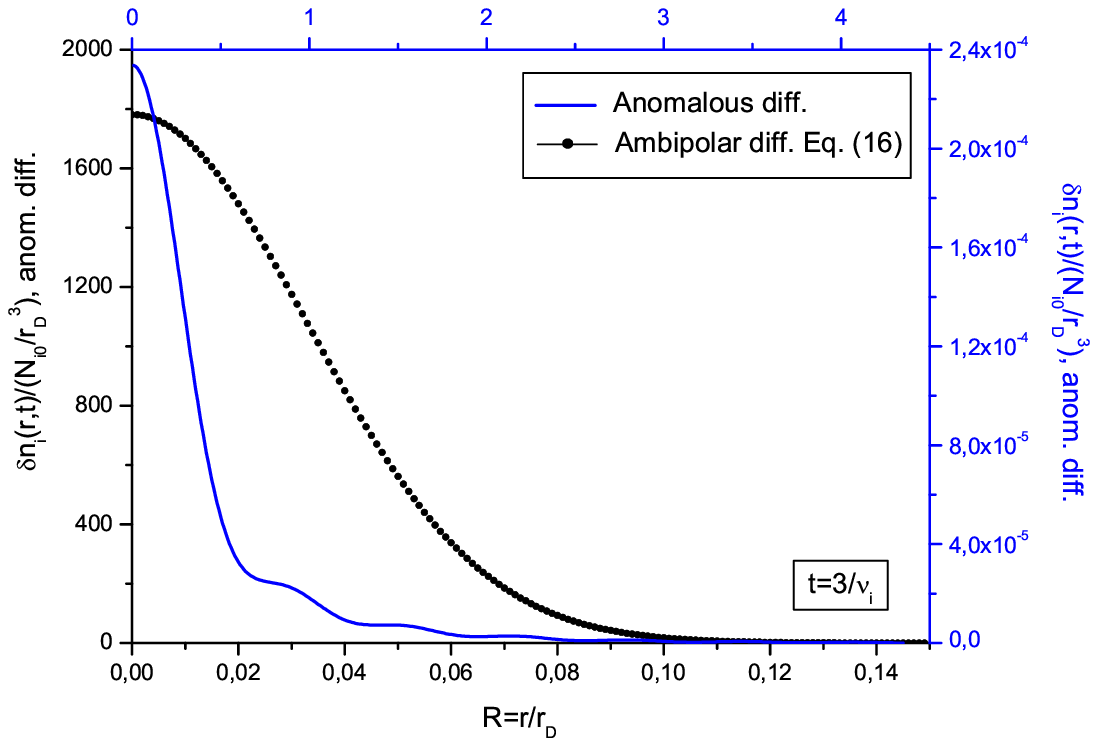}~c)}
\subfigure
{\includegraphics[height=6.8 cm,width=7.7cm]{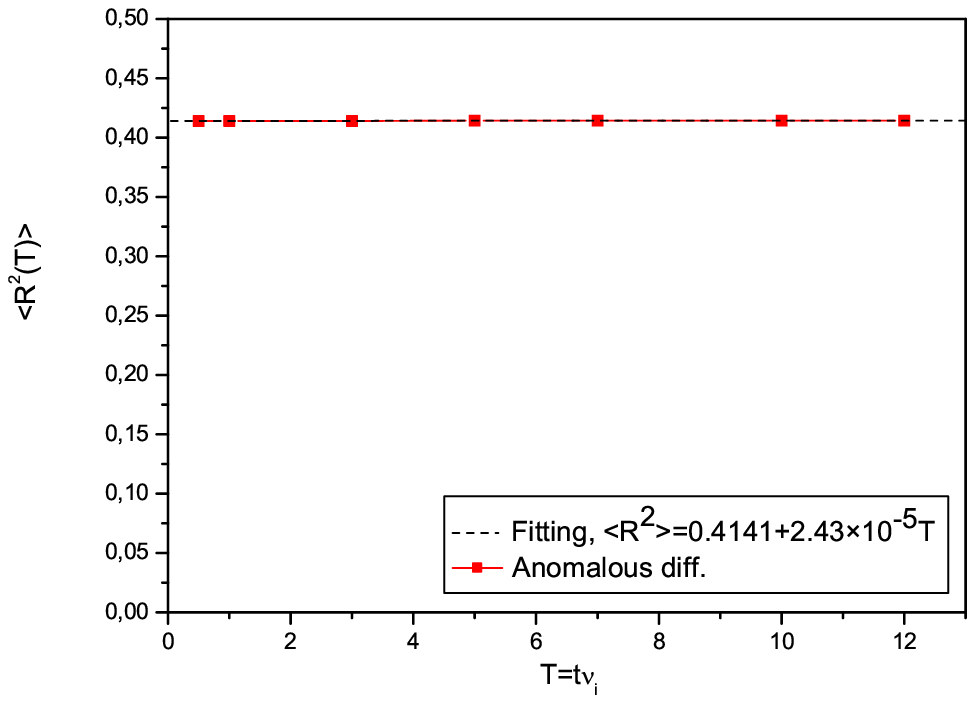}~d)}
\caption{ Comparison between the obtained the relative density perturbation relaxation for anomalous diffusion Eq. (\ref{701}) and normal ambipolar diffusion Eq. (\ref{16}) at the different time moments $T=t\nu_{i}$  but fixed Langmuir frequency $\omega_{Li}=1.6 \cdot 10^{7} $ s$^{-1}$, $n_{i0}=1.5 \cdot 10^{8}$ cm$^{-3}$  and its variance Eq. (\ref{111}) together with the variance fit d) are shown. Here  $T_e=10$ eV, $\nu_i=1.2 \cdot 10^{9}$ s$^{-1}$, $n_0=3 \cdot 10^{16}$ cm$^{-3}$ (atom density).
  }
\label{Fig:Delta0}
\end{figure*}

\begin{figure*}
\subfigure
{\includegraphics[height=6.8 cm,width=7.7cm]{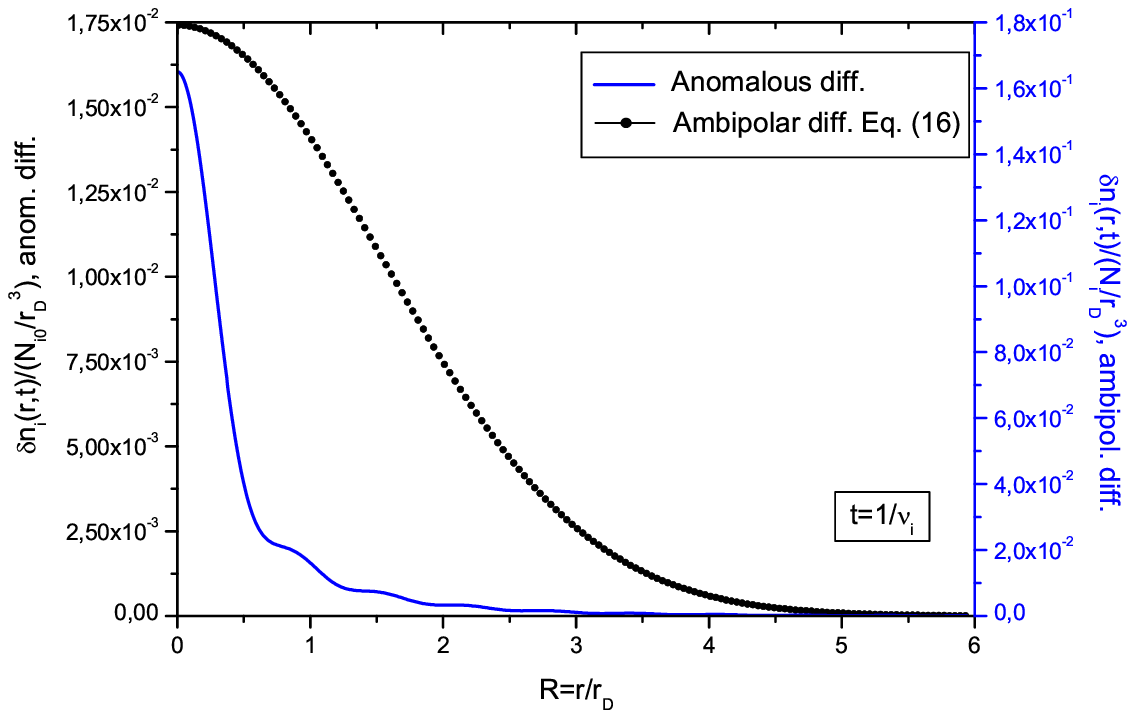}~a)}
\subfigure
{\includegraphics[height=6.8 cm,width=7.7cm]{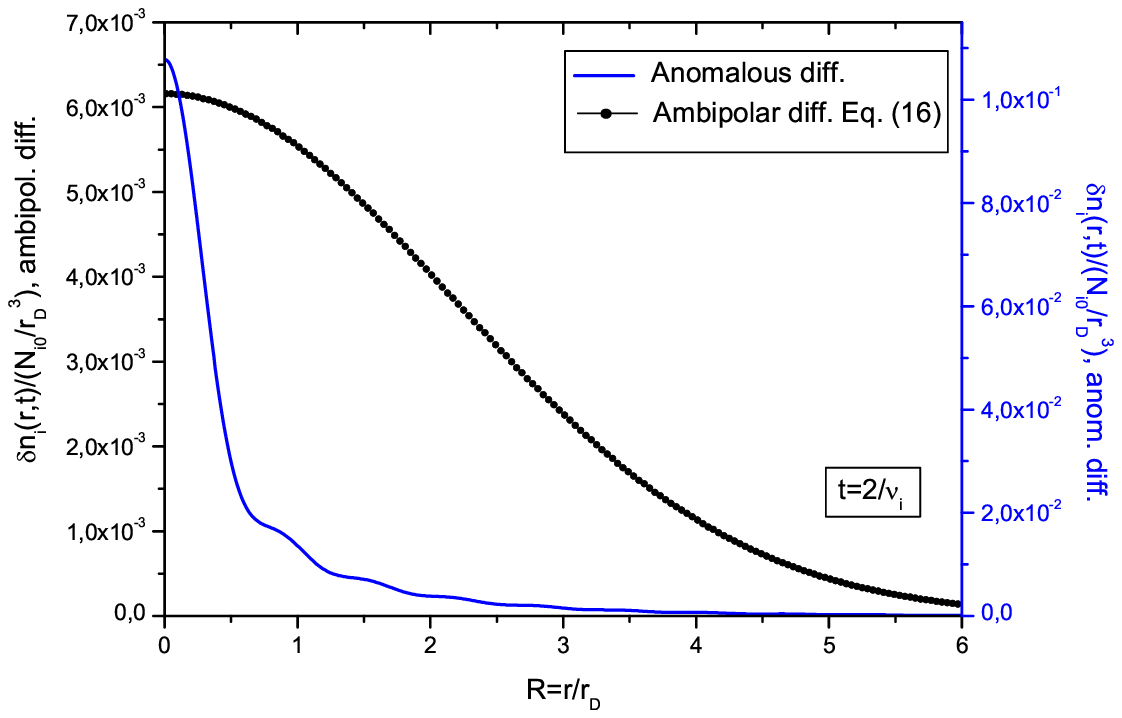}~b)}\\
\subfigure
{\includegraphics[height=6.8 cm,width=7.7cm]{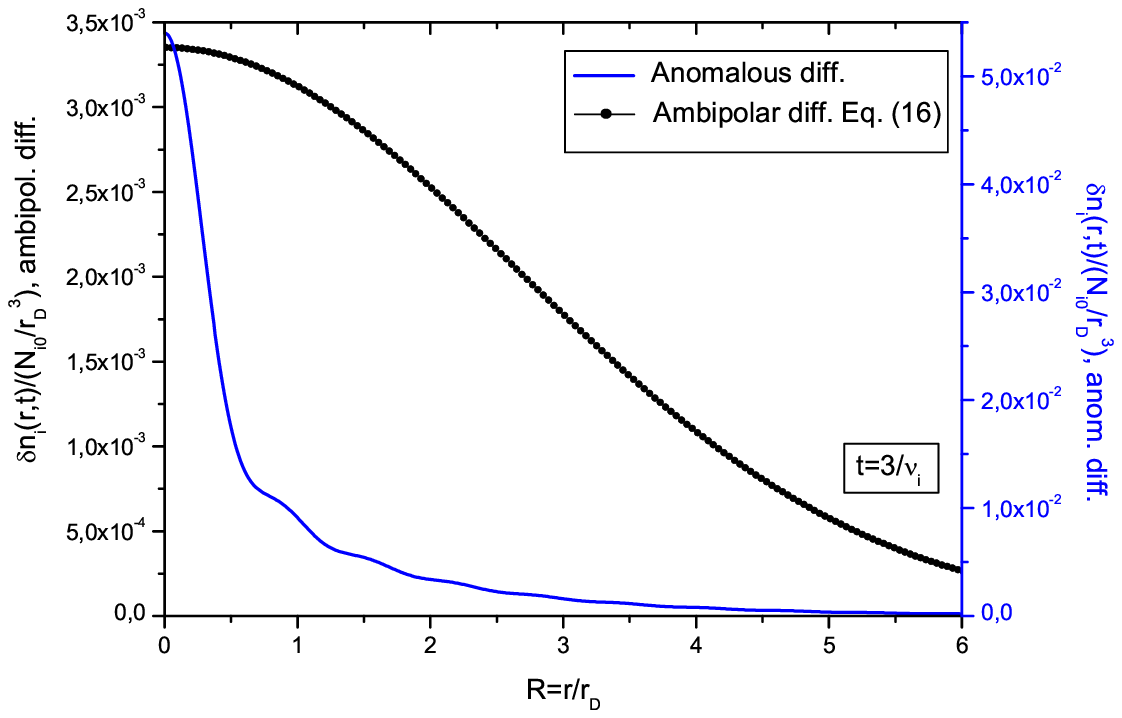}~c)}
\subfigure
{\includegraphics[height=6.8 cm,width=7.7cm]{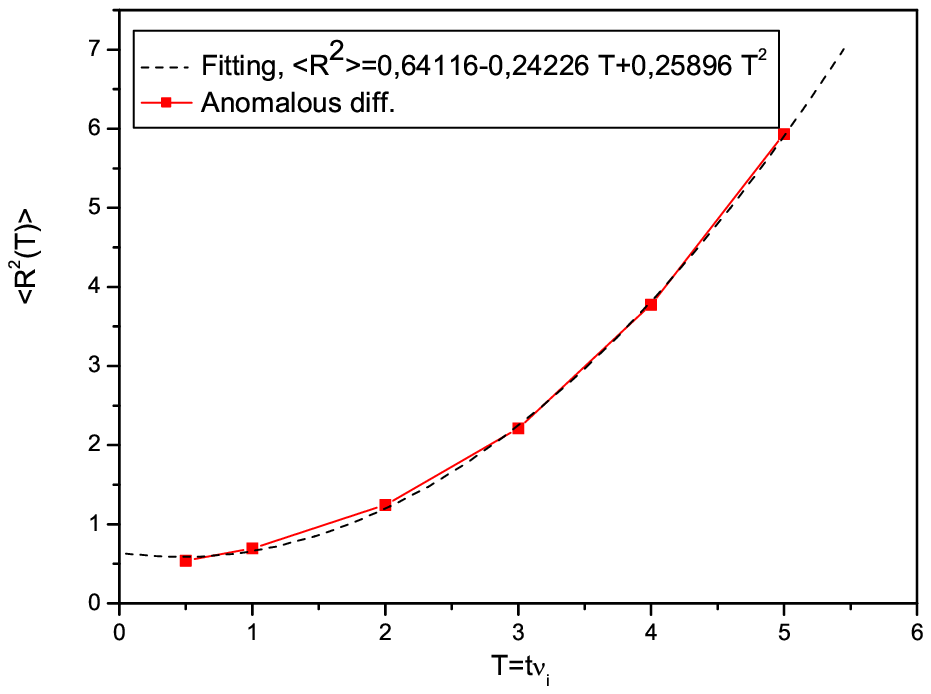}~d)}
\caption{ Comparison between the obtained the relative density perturbation relaxation for anomalous diffusion Eq. (\ref{701}) and normal ambipolar diffusion Eq. (\ref{16}) at the different time moments $T=t\nu_{i}$  but fixed Langmuir frequency $\omega_{Li}=1.3 \cdot 10^{9} s^{-1} $, $n_{i0}=10^{12}$ cm$^{-3}$ and its variance Eq. (\ref{111}) together with the variance fit d) are shown. The Hydrogen plasma parameters are the same as in the Fig. \ref{Fig:Delta0}.
}
\label{Fig:Delta1}
\end{figure*}

\begin{figure*}
\subfigure
{\includegraphics[height=6.8 cm,width=7.7cm]{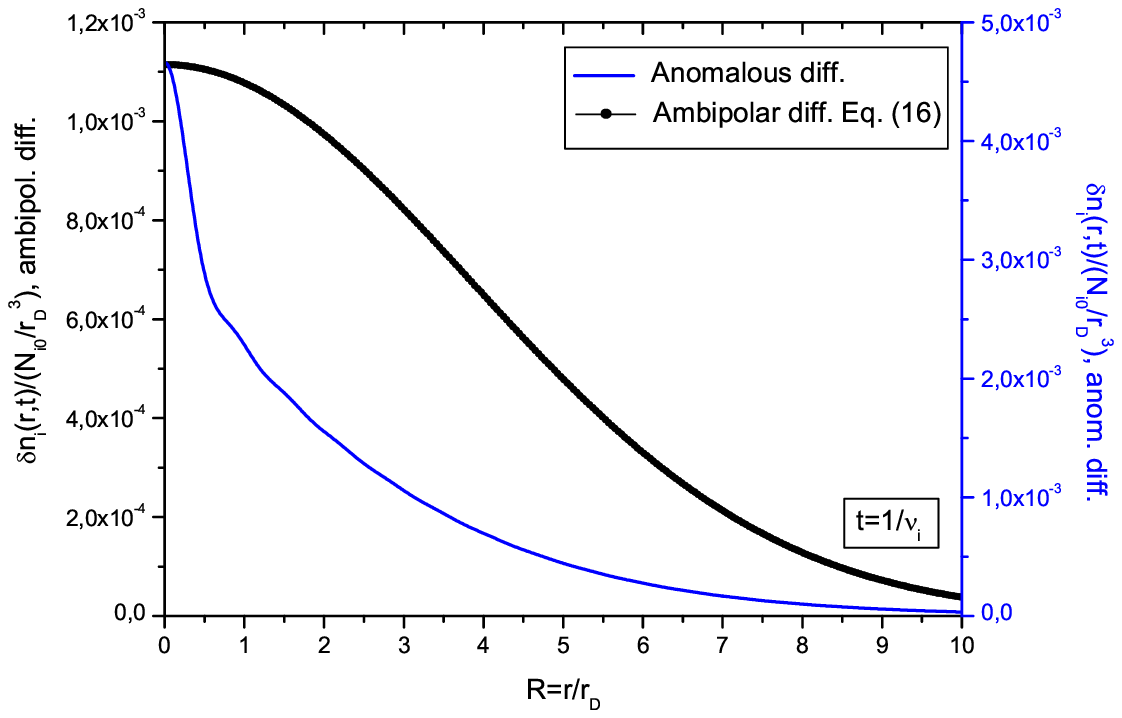}~a)}
\subfigure
{\includegraphics[height=6.8 cm,width=7.7cm]{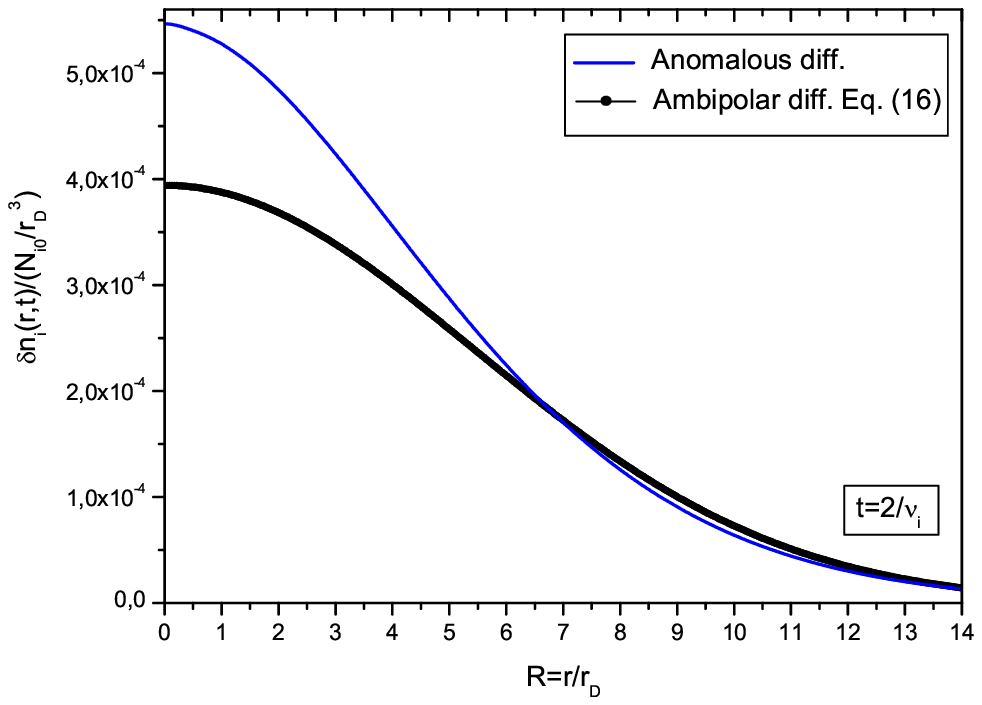}~b)}\\
\subfigure
{\includegraphics[height=6.8 cm,width=7.7cm]{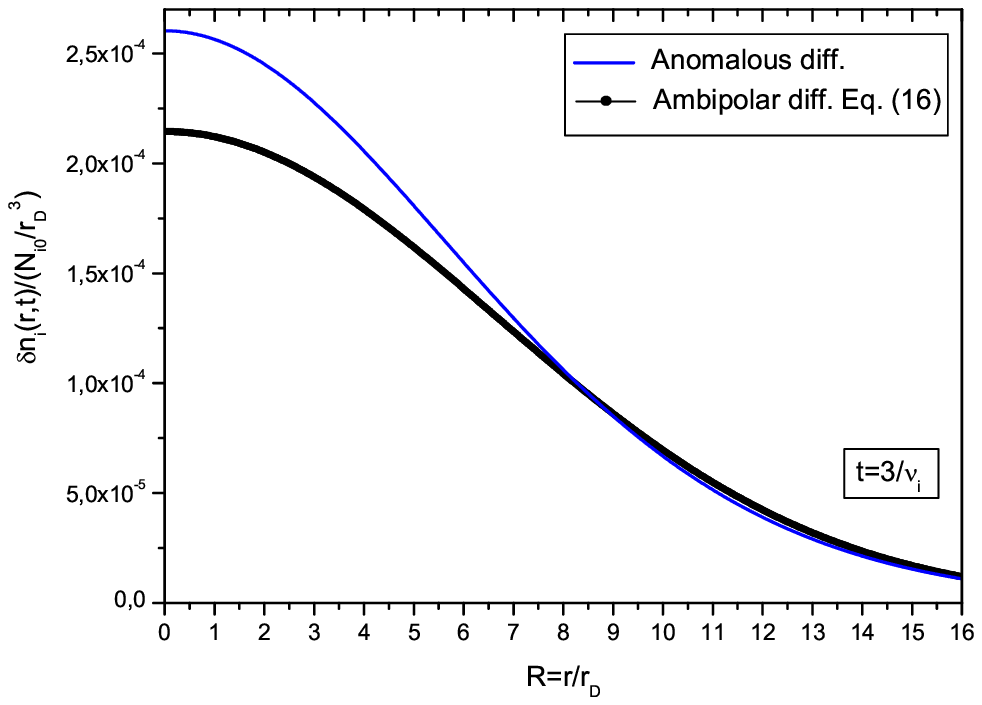}~c)}
\subfigure
{\includegraphics[height=6.8 cm,width=7.7cm]{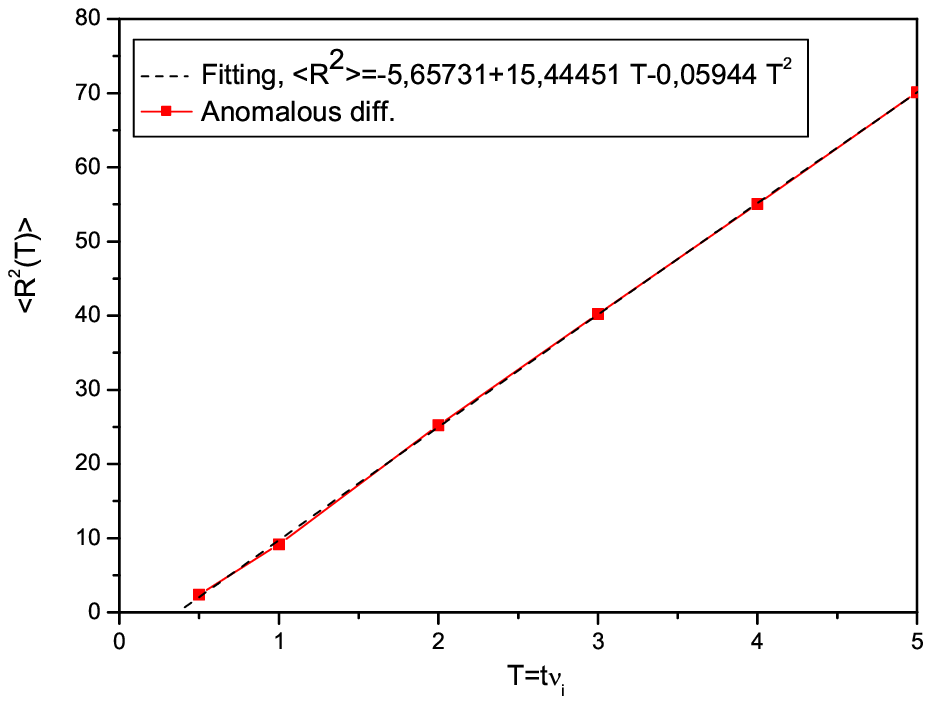}~d)}
\caption{Comparison between the obtained the relative density perturbation relaxation for anomalous diffusion Eq. (\ref{701}) and normal ambipolar diffusion Eq. (\ref{16}) at the different time moments $T=t\nu_{i}$  but fixed Langmuir frequency $\omega_{Li}=3.3 \cdot 10^{9}$ s$^{-1} $, $n_{i0}=6.25\cdot 10^{12}$ cm$^{-3}$ and its variance Eq. (\ref{111}) together with the variance fit d) are shown.  The Hydrogen plasma parameters are the same as in the Fig. \ref{Fig:Delta0}.
}
\label{Fig:Delta2}
\end{figure*}

\begin{figure*}
\subfigure
{\includegraphics[height=6.8 cm,width=7.7cm]{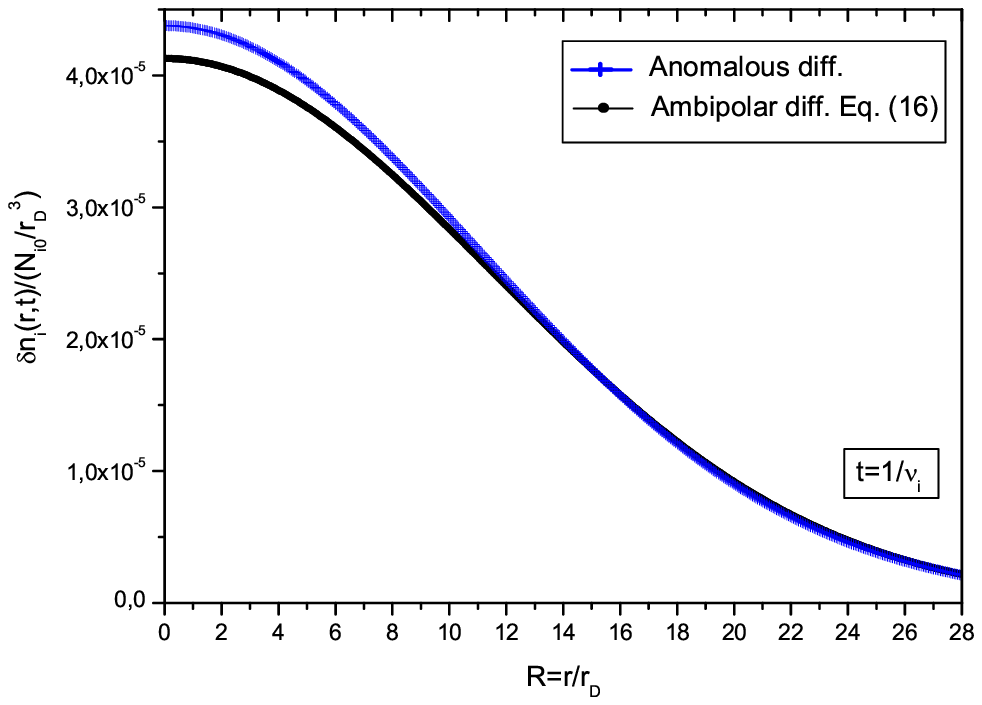}~a)}
\subfigure
{\includegraphics[height=6.8 cm,width=7.7cm]{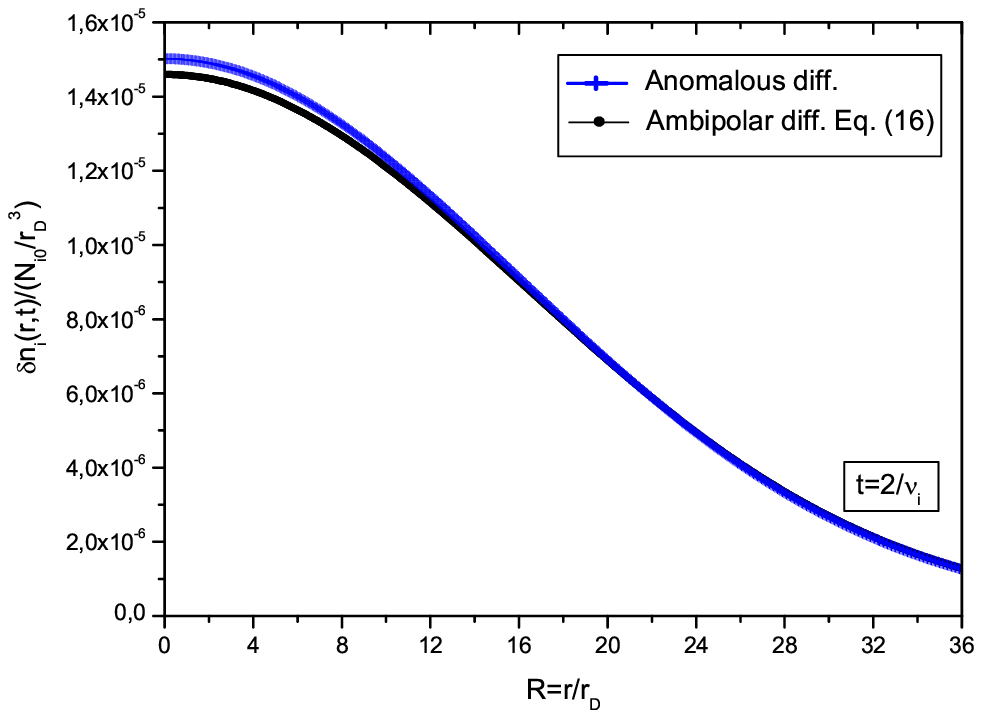}~b)}\\
\subfigure
{\includegraphics[height=6.8 cm,width=7.7cm]{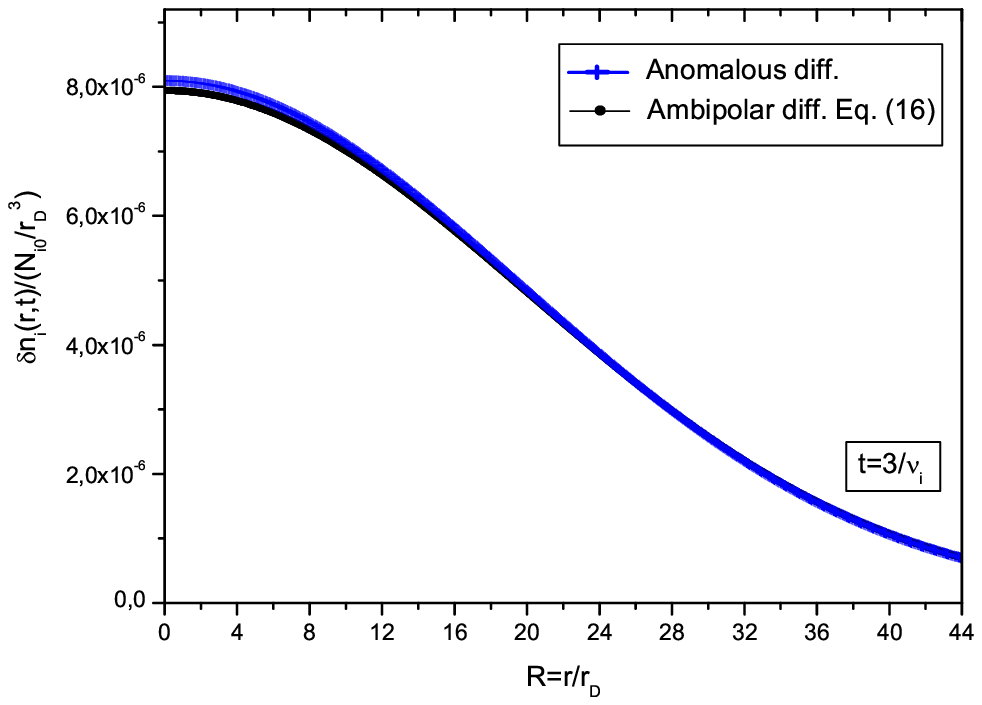}~c)}
\subfigure
{\includegraphics[height=6.8 cm,width=7.7cm]{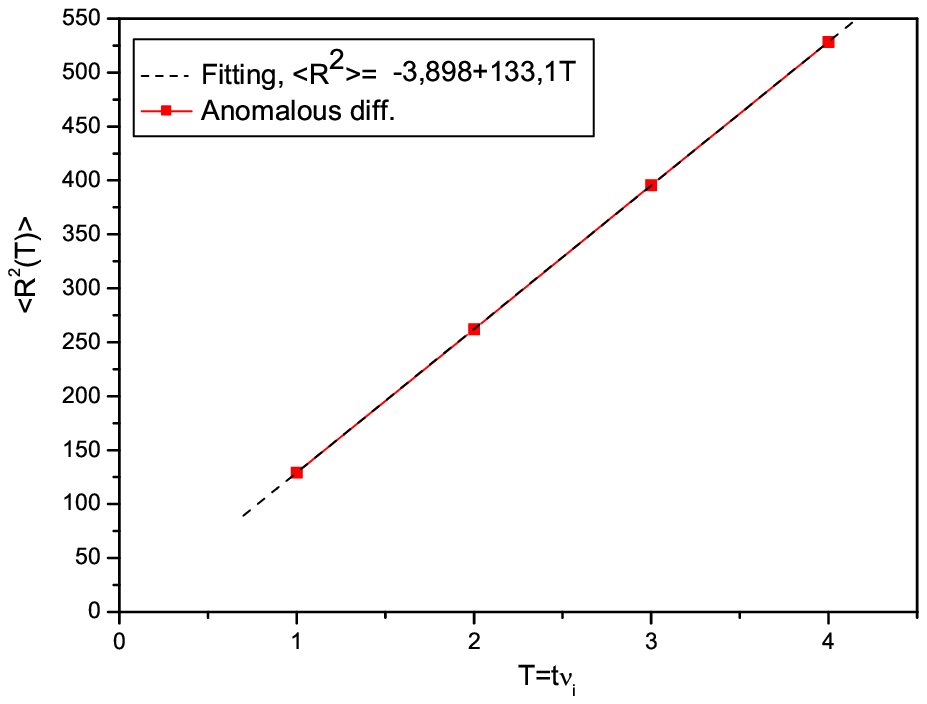}~d)}
\caption{Comparison between the obtained the relative density perturbation relaxation for anomalous diffusion Eq. (\ref{701}) and normal ambipolar diffusion Eq. (\ref{16}) at the different time moments $T=t\nu_{i}$  but fixed Langmuir frequency $\omega_{Li}=9.7 \cdot 10^{9}$ s$^{-1} $, $n_{i0}=5.6 \cdot 10^{13}$ cm$^{-3}$ and its variance Eq. (\ref{111}) together with the variance fit d) are shown.  The Hydrogen plasma parameters are the same as in the Fig. \ref{Fig:Delta0}
}
\label{Fig:Delta3}
\end{figure*}

In the Figure \ref{Fig:Anom}~a) the 3 dimensional view of the relative value of density perturbation in dependence on time $T$ and absolute value of the radius vector $R$ is shown. Namely, Fig. \ref{Fig:Anom}~a) represents the anomalous relaxation as the combined effect Eq. (\ref{23}) of the Maxwellian relaxation Eq. (\ref{19}) and ambipolar diffusive relaxation Eq. (\ref{21}). In Fig. \ref{Fig:Anom}~b) the comparison among the mentioned relaxations  and the localized ion density perturbation relaxation Eq. (\ref{16}) at $t=1/\omega_{Li}$ is shown. \\
\indent As one can easily see the behaviour of the  Maxwellian relaxation Eq. (\ref{19}), ambipolar diffusive relaxation Eq. (\ref{21}), and  the anomalous diffusion Eq. (\ref{16}) is oscillatory and fades down with the time as the $\exp(-\alpha t)\cos(\beta R)$ function. In a case for the localized ion density perturbation relaxation Eq. (\ref{16}) the perturbations fades down as the $\exp(-\gamma R^2/t)/t^{3/2}$.\\
\indent At the given oscillatory initial perturbation (\ref{17}) no normal diffusion occurs: the initial perturbation gets faded down  with the time evolution keeping its initial spatial functional form  - $\cos(\beta R)$, i.e. only the amplitude decreases, and in this way revealing the anomalous nature of the diffusion in nonisothermal plasma. This behaviour is peculiar for the Maxwellian relaxation (\ref{14}) independent on the type of initial condition. However, when we take another initial condition ($\delta(\vec r)$ - functional) for the case of diffusive relaxation (\ref{15}) then we obtain the normal ambipolar diffusion (\ref{16}). \\
\indent In the Figures \ref{Fig:Delta0}, \ref{Fig:Delta1}, \ref{Fig:Delta2}, and \ref{Fig:Delta3} comparison between the obtained relative density perturbation relaxation for anomalous diffusion Eq. (\ref{701}) with $\delta(\vec r)$ as initial perturbation (\ref{700}) and normal ambipolar diffusion Eq. (\ref{16}) and corresponding  variances at the different ion Langmuir frequencies and different time moments are shown.  In the Figs. \ref{Fig:Delta0}, \ref{Fig:Delta1}  the difference between the obtained curves for anomalous diffusion and the normal ambiplar diffusion is quite significant, whereas with an increase of either ion Langmuir frequency or time the anomalous diffusion gets to turn into the normal ambipolar diffusion. This is also demonstrated in the correspoding Figs. ~d) where dependence of the variance on time and the fitting curves for the corresponding different Langmuir frequencies is shown. The variance was estimated using Eq. (\ref{111}). As we can see at the lowest frequency there is almost no diffusion Fig. \ref{Fig:Delta0}~d), the variance does not depend on time. However, with further increase of the ion frequency the diffusion becomes anomalous $<r^2(t)>\neq \alpha t$ ($\alpha$=Const), whereas when the frequency and time moments increase the diffusion starts to turn into the normal:  $<r^2(t)>=\alpha t$  (\ref{112}).

\begin{acknowledgments}
One of the authors S.P. Sadykova would like to thank her coauthor A. A. Rukhadze for his fruitful contribution to the work and discussions on the presented topic. S.P. Sadykova would like to express her gratitude to her father P. S. Sadykov for his financial support of the work and for being all the way the great moral support for her.
\end{acknowledgments}

\appendix

\nocite{*}

\end{document}